\documentclass[suppldata]{interact}

\usepackage{epstopdf}
\usepackage{verbatim}
\usepackage[caption=false]{subfig}
\usepackage{color}
\usepackage[longnamesfirst,sort]{natbib}
\bibpunct[, ]{(}{)}{;}{a}{,}{,}

\theoremstyle{plain}

\theoremstyle{definition}

\theoremstyle{remark}

\begin{document}


\title{A Statistical Exploration of Duckworth-Lewis Method Using Bayesian Inference }

\author{
\name{Indrabati Bhattacharya\textsuperscript{a} Rahul Ghosal\textsuperscript{a} and Sujit Ghosh\textsuperscript{a}}
\affil{\textsuperscript{a} Department of Staistics, North Carolina State University}
}

\maketitle

\begin{abstract}
Duckworth-Lewis (D/L) method is the incumbent rain rule used to decide the result of a limited overs cricket match should it not be able to reach its natural conclusion. Duckworth and Lewis (1998) devised a two factor relationship between the numbers of overs a team had remaining and the number of wickets they had lost in order to quantify the percentage resources a team has at any stage of the match. As number of remaining overs decrease and lost wickets increase the resources are expected to decrease. The resource table which is still being used by ICC (International Cricket Council) for 50 overs cricket match suffers from lack of monotonicity both in numbers of overs left and number of wickets lost. We apply Bayesian inference to build a resource table which overcomes the non monotonicity problem of the current D/L resource table and show that it gives better prediction for teams in first innings score and hence it is more suitable for using in rain affected matches.\par
\end{abstract}


\section{Introduction}
A cricket match is played between two teams each consisting of 11 players. In a 50 over cricket match also known as One Day International (\textbf{ODI}), each team has by default a maximum of 50 overs to score their runs. In an uninterrupted match the team batting second wins the match if they score more runs than the team batting first. A cricket match  is often interrupted due to weather or technological faults (rain, bad light, etc). \textbf{Duckworth-Lewis} method is the incumbent rain rule used to decide the result of a limited overs cricket match in case of an interruption.\par
The Duckworth-Lewis (D/L) method was developed by Frank Duckworth and Anthony Lewis in 1998.
as a fair way to decide the result of an interrupted game of one day cricket and it has
been universally adopted and used by International Cricket Council (ICC) since 2001.
Prior to the adoption of the Duckworth Lewis method, the resetting of targets in interrupted one day cricket matches used to be carried out via adhoc tournament specific methods or using run
rates.\par
For example, in Cricket world cup 1992 semifinal between England and South Africa, rain come down
with South Africa needing 22 runs off the final 13 balls. The game couldn't be extended because of television demands, so when the play resumed, the umpires had shortened the game to leave South Africa with just one ball . A new rule introduced for that World Cup meant the target was reduced only by the amount of runs England had scored in their least productive two overs with the bat which was 1. That meant South Africa were left needing 21 runs from their final ball and they eventually lost the match . This game still serves as an example of one of the great injustice in a cricket match. Similarly the difficulty with using run rates is that targets are determined by taking the remaining overs into account, while ignoring the number of lost wickets. So this does not consider the fact that batsmen bat less aggressively and score fewer runs when more wickets have been taken. Thus it became necessary to develop a method that could overcome these specific kind of situations by revising the required target sensibly.\par
Our paper is organized as follows. In section 2, we briefly review the  Duckworth-Lewis (D/L)  method, it's advantages and problems. In section 3 we give a brief description of our data and obtain an nonparametric resource table which still suffers from non monotonicity. In section 4 we propose an alternative Bayesian inference method to build resource tables. In section 5 we give our results, compare our resource table with the Duckworth Lewis resource table and summarize our findings and in section 6 we conclude with a discussion about possible future extensions of our work.

\section{The Duckworth and Lewis Method}
\label{sec:D/L_method}
As noted by Jack Jewson, Simon French \citep{Jewson2018} Duckworth and Lewis used the fact that a batting team have 2 resources to score runs: the batsmen they have who are yet to bat (wickets remaining) and the balls they can play (overs remaining). When an interruption (Rain) occurs overs remaining, is reduced while the number of wickets left remain intact. So they devised a two factor relationship between the numbers of overs a team had remaining and the number of wickets they had lost in order to quantify the total resources a team had remaining. This was divided by the resources each team had at the start of the game, to obtain a resource percentage, which Duckworth and Lewis tabulated allowing remaining resources to be calculated for all combinations of overs remaining and wickets lost. Then a team's target is reset such that both teams have to score proportionally the same number of runs in the percentage resources available to them.\par
They used the following exponential decay model for their two factor relationship.
\begin{equation}
R(u,w) = a_w(1-e^{-b_{w}u}) \hspace{3mm}\textit{$w\in \mathcal{W}=\{0,1,2....9\}$ and $u\in \mathcal{U}=\{0,1,2,....50\}$}
\end{equation}
where $R(u,w)$ denotes average runs scored by a team when $u$ overs are available, and $w$ wickets have been lost. The parameters of this model are $a_w,b_w,(w=0,1,2,...9$). Due to confidentiality they did not reveal the parameters or how they were estimated. Here $a_w$ is the asymptotic average runs scored by the last $10-w$ wickets in hypothetical infinite overs under one day rules and $b_w$ is the decay parameter which again depends on the number of wickets that have been lost. Once they had estimated the parameters and plugged them in equation (1), they came up with a resource percentage table, outlining the percentage resources, $P(u,w)$ a team has remaining for varying u and w in the following way.
\begin{equation}
P(u,w) = R(u,w)/R(50,0)
\end{equation}
The most recent (2013) resource table used by ICC obtained using D/L method looks like Table \ref{tab:tab1}. The full table is given in appendix. 
From these resource percentages the proportion of the full resources that a team has available after a delay in their innings is calculated. Therefore if team 1 scored $S$ in $P_{1}$ percentage resources available and team 2 has $P_{2}$ percentage resources
available, then team 2's winning target, {T}, according to D/L method is 
\begin{equation}
T=
\begin{cases}
S\frac{P_{2}}{P_{1}}+1 \hspace{3.6 cm} \text{if $P_1>P_2$}\\
S+1 \hspace{4 cm} \text{if $P_1=P_2$}\\
S + G(50)(P_{2}-P_{1})+1 \hspace{.9 cm} \text{if $P_1<P_2$}\\
\end{cases}
\end{equation}
$G(50)$ is the average runs scored in the first innings by a team batting for 50 overs. Ideally we would like $P(u_1,w)\leq P(u_2,w)$ if $u_1\leq u_2$ and $P(u,w_1)\leq P(u,w_2)$ if $w_1<w_2$. 
\subsection*{Pros and Cons of D/L method}
\begin{itemize}
\item D/L method gives a reasonable and sensible target in most of the situations.
\item More importantly it is fairly easy to apply requiring just a table and a calculator.
\item The biggest problem in the D/L method is that the resource table they use for revising scores is not monotone as can be seen in Table \ref{tab:tab1} for wickets lost 8 and 9 (last two columns) and in last row. It is not sensible for resources to remain constant as the overs are decreasing or wickets are falling.
\item Since its inception there has been criticism and debate over its actual performance in many important cricket matches.
\end{itemize}
Our goal in this paper is to build a resource table using Bayesian inference which retains the essential advantages of the Duckworth Lewis method but overcomes the non monotonicity problems it induces and also gives better predictions across matches in different situations.


\section{A Motivating Data Set}
\begin{figure}[ht]
\includegraphics[width=1\linewidth , height=.8\linewidth]{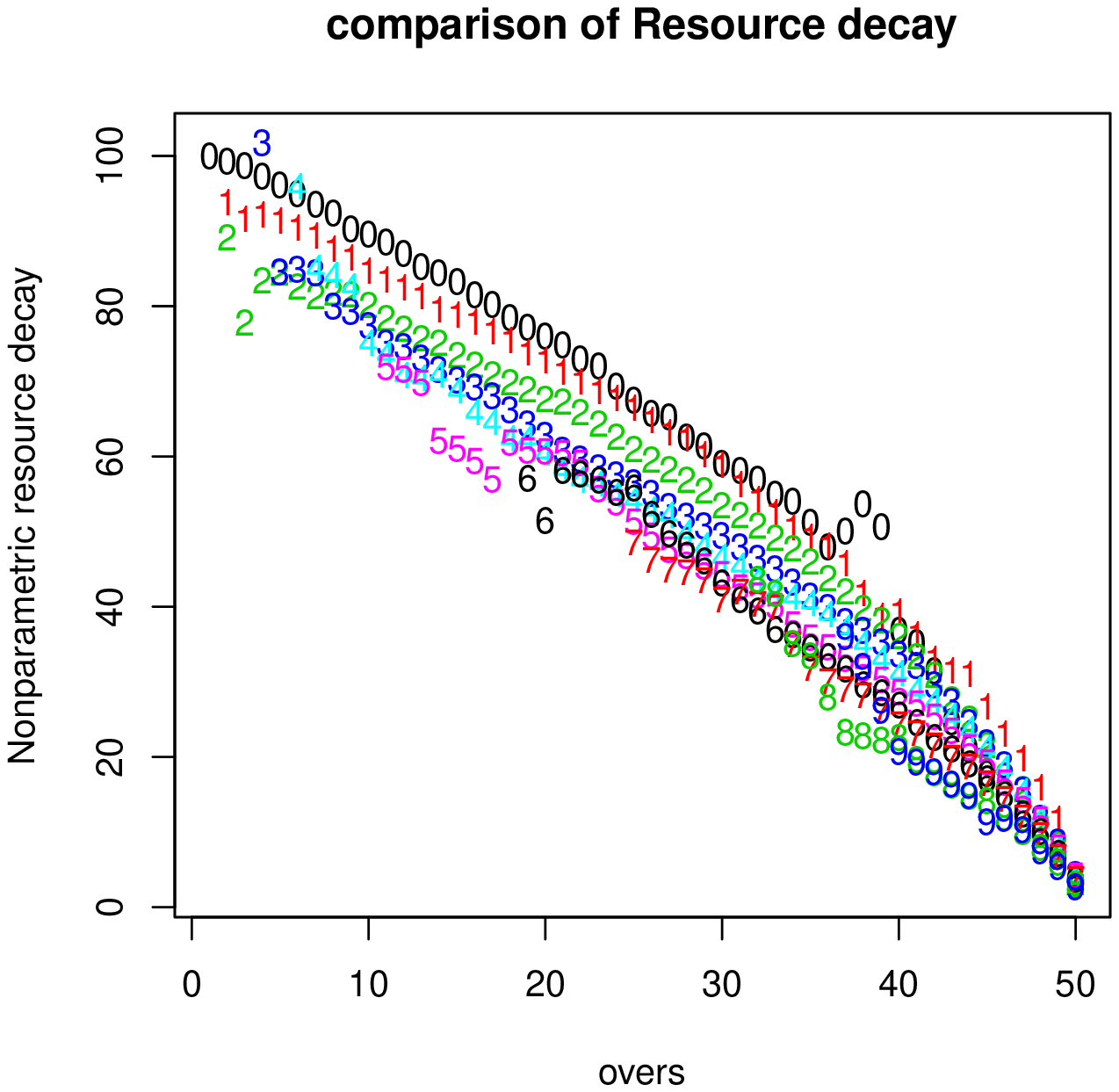}
\caption{Resource Remaining for non-parametric Method}
\label{fig:fig1}
\end{figure}

In order to illustrate our methodology we have selected to use data from all ODI matches played between 2005-2017 (downloaded from www.cricsheet.org). We have used only those 947 number of matches in which the first innings lasted for entire 50 overs. After some pre processing, we obtained the over by over runs scored and wickets lost in first innings for all these matches. In D/L method also, only the first innings data is used because Duckworth and Lewis argued that only the first innings data is relevant in producing resource percentages as teams batting second want to optimise their chance of winning and therefore does not necessarily bat in a way that optimises runs scoring. Bhattacharya, Gill and Swartz \citep{Bhat2011} pointed out that the D/L model is based on a specific (exponential) parametric form and suggested the use of non-parametric (empirical) approach to calculate the resource table. Following their method, for each match we define $R(u,w(u))$ as the run scored from the stage in first innings where $u$ overs are available and $w(u)$ wickets are lost until the end of the first innings. We calculate $R(u,w(u))$ for all values of $u$ that occurred in the first innings. The estimated resource percentage table is then calculated by averaging $R(u,w(u))$ over all matches where $w(u) = w$ and dividing by the average of $R(50,0)$ (which is the average first innings score) over all matches. Table \ref{tab:tab2} shows a part of this resource table based on the 947 matches that we have used for our illustration. The full table is given in appendix.
Just like D/L table, this non-parametric resource table suffers from the lack of monotonicity. Bhattacharya, Gill and Swartz have used isotonic regression method to overcome this issue, whereas we have taken a parametric Bayesian approach as described in the next section.
Figure \ref{fig:fig1} shows how this non-parametric model based resources decay as overs remaining decrease for different wickets. Throughout this paper in resource decay plots index $w \in \mathcal{W}$ indicates loss of $w$ wickets.

\section{Bayesian Modeling }
\label{sec:Bayes_mod}
There are several aspects of the data that motivates us to take a Bayesian inferential frame to obtain an estimate of the resource models. Besides the lack of monotonicity problem in Table \ref{tab:tab2}, we can also notice that there are many potentially missing entries in the table, which arises because there is no observable data in some of the extreme cases even when thousands of matches are considered in a data set. Instead of throwing out those columns or rows that have missing entries, we use the Bayesian inferential framework that provides a natural way of imputing the missing entries using the so-called posterior predictive distribution once a full hierarchical model is specified. We adopt the following nonlinear regression model:
\begin{equation}
\label{eq:lik}
\bar{R}(u,w) \sim N(m(u,w;\theta),\sigma^2/n_{uw}), \quad u \in \mathcal{U},\ w \in \mathcal{W},
\end{equation}
where $\bar{R}(u,w)$ is the sample average of runs scored by a team among the total number of matches considered in the data set and $m(u,w;\theta)$ is the corresponding modeled population average of runs scored by a team when a large number of games are taken into consideration and $\theta$ denotes a vector of parameters to be specified later in our model. As $R(u,w)$ is not observed for each of the match (in the sample), the average is taken over all those matches, denoted by $n_{uw}$, over which the sample mean $\bar{R}(u,w)$ is calculated. If there is no observation for $R(u,w)$ across all of the matches sampled, we treat the values of $\bar{R}(u,w)$ missing and assign $n_{uw}=1$. For example, in our application to a case study we considered a sample of about 1000 matches in the motivating data set presented in Section 3 and for this data set, $n_{uw}$ ranged between $0$ and $947$ across 947 matches that we sampled and about $26.8$\% values were missing. One of primary advantage of adopting a Bayesian inference framework is to use its built-in option of using the posterior predictive distribution to impute the missing values given the observed values within the so-called missing at random (MAR) assumption \citep{rubin1976inference}

Borrowing  ideas from the popular and current used D/L method  and also based on the plot of resource decay in Figure 2, we decided upon to adopt an exponential decay model for the mean function. The mean function $m(u,w;\theta)$ depends on two parameters, $a_w$ and $b_w$, which themselves depend on $w \in \mathcal{W}$, the number of wickets lost at that time (when $u$ overs remain). Therefore $m(u,w;\theta)$ will be of the form
\begin{gather}
\label{eq:mean}
m(u,w;\theta) = a_w(1-e^{-b_w u})\\
\mbox{ and }\;\theta = \{(a_w, b_w); w\in \mathcal{W}\}. \notag
\end{gather}
While considering a prior distribution for the vector of parameters $\theta$, we need to ensure (with probability 1) that the following conditions are satisfied by the prior distribution
\begin{eqnarray}
\label{eq:nec.cond}
m(u_1,w) &<& m(u_2,w)\;\;\mbox{if}\;u_1< u_2,\ u_1,u_2 \in \mathcal{U},\ w \in \mathcal{W}\\
m(u,w) &>& m(u,w+1)\;\;\forall u \in \mathcal{U},\ w \in \mathcal{W}. \notag
\end{eqnarray}
This can be done by the following prior specifications on the parameters which also leads to relatively non-informative prior (meaning that the posterior inferences will not be sensitive to the choice of this prior). Fix $A_0$ and $B_0$ large enough such that $R(50,0)<A_0$. Initialize with $a_0\sim U(0,A_0)$ and $b_0\sim U(0,B_0)$. Then given $(a_0,b_0)$ generate for $w=0,1,...8$   
\begin{eqnarray}
\label{eq:prior}
\ a_{w+1}| \sigma^2,a_w,b_w &\sim& U(0,a_w) \\
\ b_{w+1}| \sigma^2,a_{w+1},b_w,a_w &\sim& U\left(0,\frac{a_wb_w}{a_{w+1}}\right) \notag\\
1/\sigma^2 &\sim& Ga(a,b), \notag
\end{eqnarray}
where $U(c, d)$ denotes a uniform distribution over the interval $(c, d)$ and $Ga(a, b)$ denotes the Gamma distribution with mean $a/b$ and variance $a/b^2$. Clearly, the above sequence of conditional distributions in (\ref{eq:prior}) defines a valid joint prior distribution for the entire parameter vector $\theta$ as given in (\ref{eq:mean}). It can be shown (see Appendix B for a detailed proof) that above prior distribution specified by the sequence of conditionals in (\ref{eq:prior}) ensures (with probability 1) that the necessary conditions for monotonicity given in (\ref{eq:nec.cond}) is satisfied. 

An alternative way of specifying a prior distribution on $\theta$ can be obtained by a re-parametrization $c_w=a_wb_w$ and $a_w$ as defined earlier, we have $m(u,w;\eta)= a_w(1-e^{-\frac{c_w}{a_w} u})$, where $\eta=\{(a_w, c_w); w\in\mathcal{W}\}$. Again as before we initialize with $a_0\sim U(0,A_0), c_0\sim U(0,C_0)$ where $C_0$ is a fixed large constant, we specify the conditionals as
\begin{eqnarray*}
\ a_{w+1}| \sigma^2,a_w,c_w &\sim& U(0,a_w) \\
\ c_{w+1}| \sigma^2,a_{w+1},c_w &\sim& U(0,c_w)\\
1/\sigma^2 &\sim& Ga(a,b).
\end{eqnarray*}
Again it follows (see Appendix B) that the above set of conditionals creates a joint prior distribution for $\eta$ which ensures (with probability 1) that (\ref{eq:nec.cond}) is satisfied. As there is no clear advantage of using one prior over the another, for our numerical illustrations, we have used the prior distribution as specified by the conditional in (\ref{eq:prior}). 

Once a prior distribution is obtained we use (full) likelihood function based on the sampling distribution specified in (\ref{eq:lik}) to obtain the posterior distribution. The likelihood function is then given by
$$L(\theta,\sigma^2)=\left(\frac{1}{\sigma/\sqrt{n_{uw}}}\right)^{500} \mathrm{exp}\big\{-\frac{1}{2(\sigma^2/n_{uw})}\sum_{u=1}^{50}\sum_{w=0}^{9}\left(R(u,w)-m(u,a_w,b_w)\right)^2\big\},$$
where $\theta$ is as defined in (\ref{eq:mean}). Also the joint prior density of $\theta$ is obtained by multiplying the following conditional densities:
\begin{gather*}
\pi(a_0)=\frac{1}{A_0}\mathbb{I}(0<a_0<A_0)\\
\pi(b_0)=\frac{1}{B_0}\mathbb{I}(0<b_0<B_0)\\
\pi(a_{w+1}|\sigma^2,a_w,b_w)=\frac{1}{a_w}\mathbb{I}(0<a_{w+1}<a_w)\;\; \mbox{for}\;w \in \mathcal{W}\\
\pi(b_{w+1}|\sigma^2,a_{w+1},a_w,b_w)=\frac{1}{\frac{a_wb_w}{a_{w+1}}}\mathbb{I}(0<b_{w+1}<\frac{a_wb_w}{a_{w+1}}) \;\; \mbox{for}\;w \in \mathcal{W}\\
\pi(\sigma^2)=\frac{b^a}{\Gamma(a)}(\sigma^2)^{-(a+1)}e^{-{b\over \sigma^2}}
\end{gather*}
The choice of hyper-parameters, $A_0$, $B_0$, $a$ and $b$ will be such that the prior is not sensitive to the posterior inference. In all of our numerical illustrations, we have chosen $A_0=2000$ and $B_0=100$ and $a=b=0.1$ which results into so-called vague prior with extremely large variances compared to the posterior variances. The analytical expression for the joint posterior distribution of $(\theta, \sigma^2)$ is rather long and complicated form due to various (nonlinear) order restrictions among the parameters $a_w$ and $b_w$. For this reason we use Monte Carlo sampling based methods to draw approximate samples from the resulting posterior distribution using the so-called Markov-Chain-Monte-Carlo (MCMC) methods and in particular, we use Gibbs Sampling (to obtain samples from the full conditional distribution of $\sigma^2$ given $\theta$ and observed data values) and Metropolis-Hastings (MH) algorithm \citep{Hastings1970monte} to obtain the samples from the full conditional distributions of $a_w$ and $b_w$'s given rest of the parameters and data values. The full conditional distribution of $\sigma^2$ can be shown to be an inverse gamma distribution, while the conditional posterior densities of the other parameters do not have a standard closed form. Therefore, we use the slice sampling within the MH algorithm to sample the components of $\theta$ given the samples of other components. In particular, we have used the freely available software JAGS \citep{plummer2003jags} for generating posterior samples from the above mentioned model which identifies the target density as non log concave and uses a slice sampling algorithm for running the full cycle of MCMC chains. Finally, for missing values of the $\bar{R}(u,w)$ it easily follows that the conditional posterior predictive distribution of missing values given the sampled parameter values and observed values is a normal distribution and hence can be easily sampled and used as imputed values. A snippet of the JAGS model code is given in the Appendix C for the first set of prior distributions.

\section{Numerical Results}
\subsection{Bayes Resource Tables and Decay Graphs}
\begin{figure}[ht]
\hspace*{-5em}
\includegraphics[width=1.3\linewidth , height=1.4\linewidth]{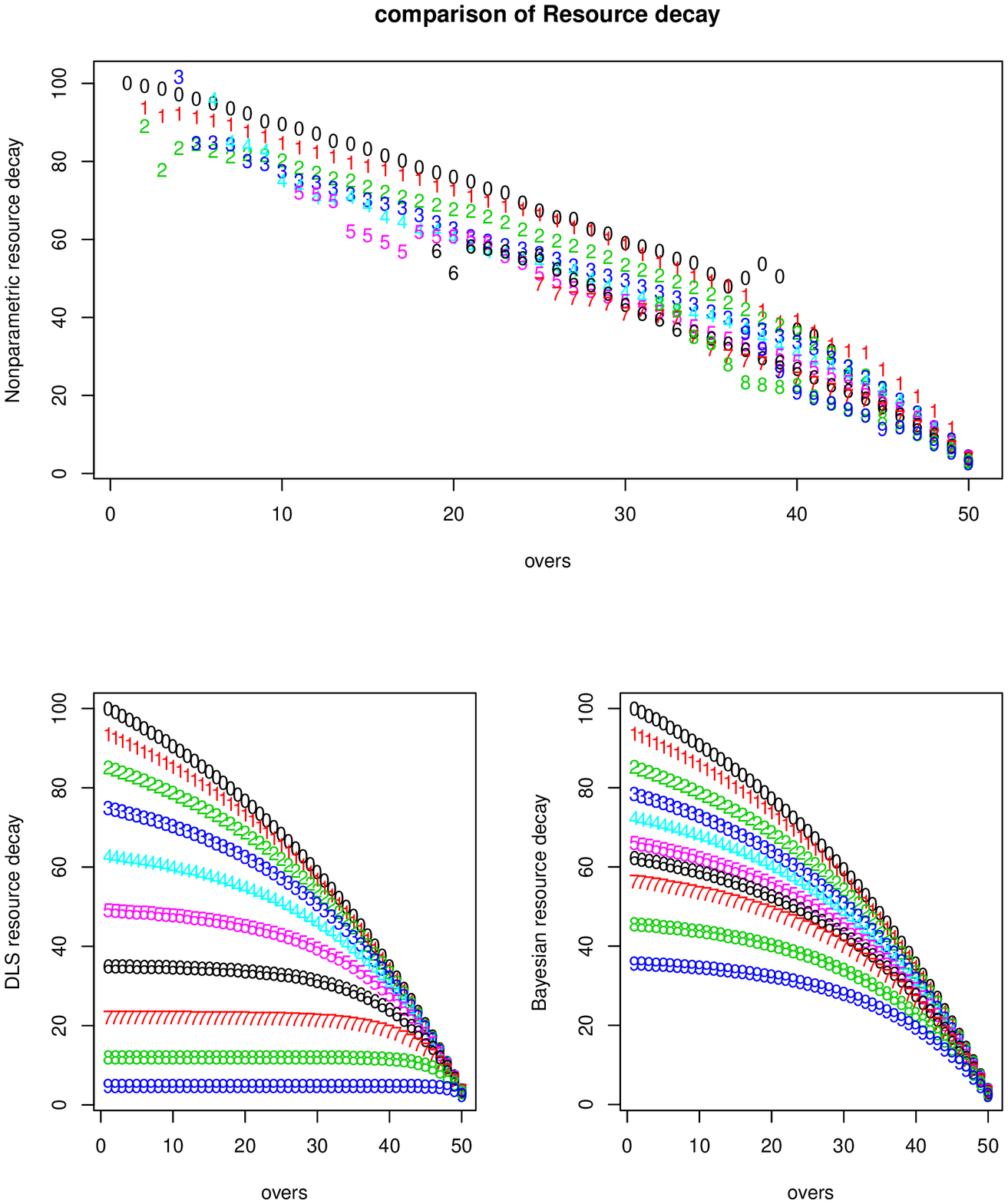}
\caption{ Comparison of Resource Decay}
\label{fig:fig2}
\end{figure}

We fitted the model in terms of runs obtainable instead of percentage resources as was originally done by D/L and since there is a functional correspondence between percentage resources remaining and runs obtainable, resources are simply extracted by dividing the runs obtainable by the runs the model expects a team to score when $100\%$ of their resources are available. We used JAGS for generating posterior samples from the above mentioned model. We used a `burn-in' of 20000 samples and used the subsequent 30000 samples for posterior inference.The posterior medians of the parameters have been used as the estimates as robust alternative to posterior means. A snapshot of the resource tables generated using these estimates for our Bayesian  Model is  shown in Table \ref{tab:tab3}. The Full resource tables are provided in appendices.
We can notice that both these tables are monotone across rows and columns. Figure \ref{fig:fig2} shows a comparison of how the resources change as overs remaining decrease for for different wickets for the D/L method, non-parametric model and our Bayesian model. The Bayesian models seem to capture the resource decay more accurately.

\subsection{Score Prediction}
\begin{figure}[ht]
\hspace*{-4em}
\includegraphics[width=1.3\linewidth , height=1.4\linewidth]{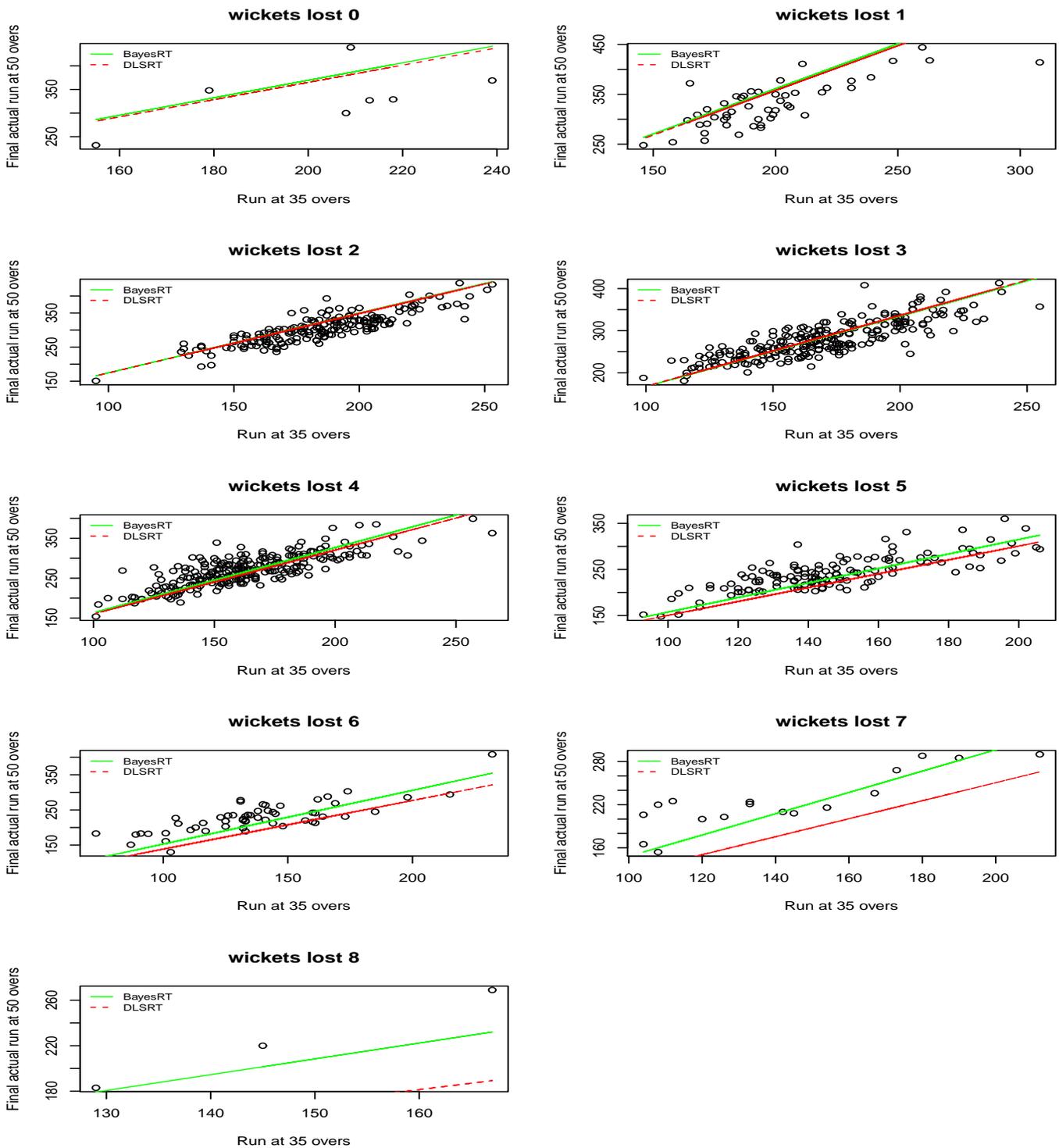}
\caption{Comparison of Predictions at the end of 35th over}
\label{fig:fig3}
\end{figure}

\begin{figure}[ht]
\includegraphics[height=9cm,width=14cm]{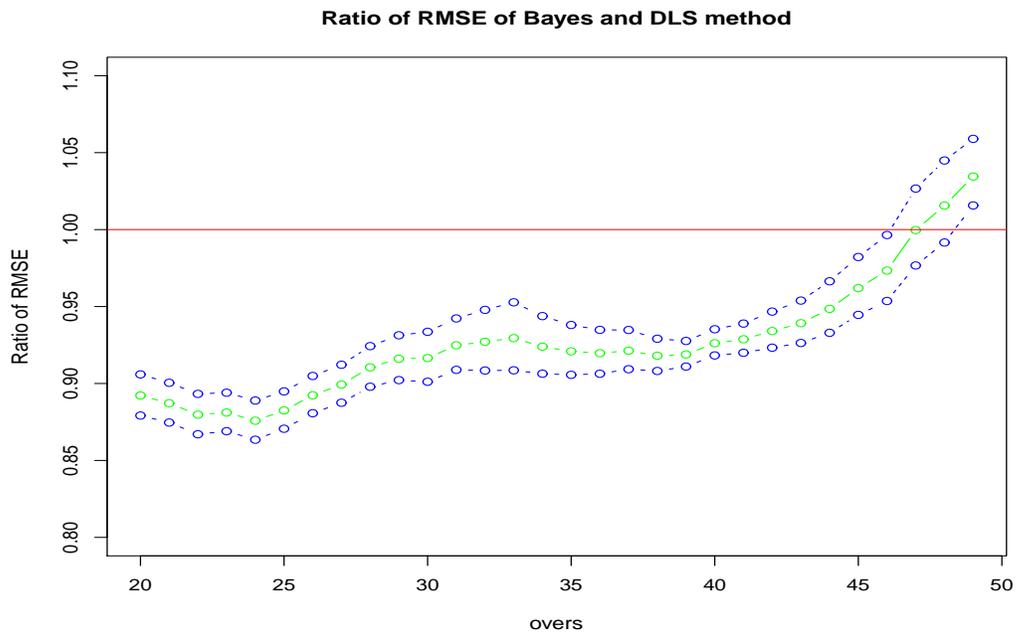}
\caption{Comparison of Ratio of Residual mean Sum of Squares of Bayes method and DLS method}
\label{fig:fig4}
\end{figure}

\begin{figure}[ht]
\includegraphics[height=9cm,width=14cm]{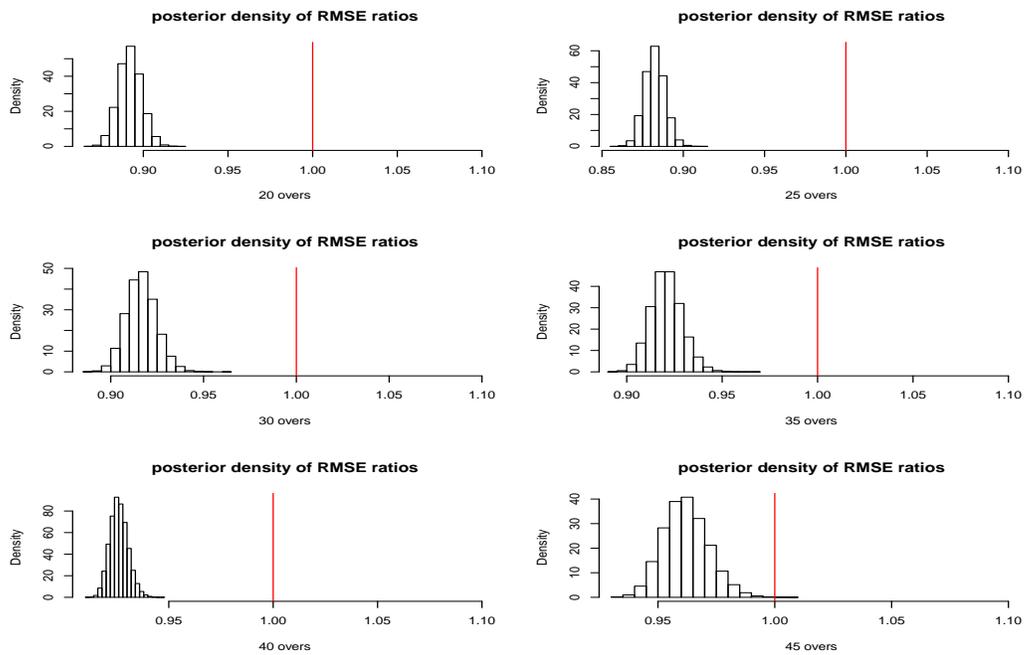}
\caption{Posterior density Ratio of Residual mean Sum of Squares of Bayes method and DLS method}
\label{fig:fig5}
\end{figure}

To see how good our models are in predicting scores, we split the 1st innings of each match at a point where number of overs left is say, $u$ and number of wickets lost is, say, $w$ and from that point we try to predict the final score using our Bayesian resource tables based on equation (\ref{eq:mean}). For a match that has been cut at the end of $(50-u)th$ over, we define a Residual sum of Squares measure as below:
\begin{equation}
RSS_u = \sum_{w=0}^{w=9} \sum_{i=1}^{n_{u,w}}(R^{A}_{u,w,i} - R^{P}_{u,w,i})^2,
\end{equation}
where $R^{A}_{u,w,i}$ is the actual final score for the $i$th match where the 1st innings has been split at $u$ overs left and $w$ wickets lost and $R^{P}_{u,w,i}$ is the predicted final score for the same. $n_{u,w}$ is the number of such matches where the team has lost $w$ wickets at $u$ overs remaining. In Figure \ref{fig:fig3}, we split each match at 35th over and for each $w$ ranging from $0$ to $9$, we plot the final scores at the end of 50 overs against runs at the end of 35th over with predictions from all the methods overlaid on them. The Bayesian model seem to give more accurate predictions of scores across all wickets.  Figure \ref{fig:fig4} shows the ratio of square root $Rss_u$'s for Bayesian method to the D/L method where $u$ (overs left) ranges from $30$ to $1$. As can be seen our Bayesian method gives smaller residual sum of squares in majority of the scenario and their difference is  statistically significant in majority of the portions of these overs. Figure \ref{fig:fig5} shows the posterior density of the ratios for match cut off at $20,25,30...45$ over marks. Again The Bayesian model gives better prediction of match scores as evident from these figures.
\section{Discussion and Future Work}
We have developed an alternative method to build resource tables using Bayesian methodology which not only removes the non-monotonicity issue present in the D/L resource table but also capture the exponential decay relationship more accurately.
We have shown the Bayesian method also gives better score predictions and performs better in terms of Residual Sum of Squares than the D/L method specially when the match is interrupted in situations where there are lots of overs left. Under the MAR assumption, the proposed Bayesian model provides a natural method to carry out imputations using the posterior predictive distributions which is an advantage over many existing methods (e.g., compared to the non-parametric method). Our method is broadly applicable in the sense that it is not restricted to only 50-overs cricket match interruption problem and can be applied many similar sports events. More generally, the model can be used to estimate the nonlinear mean mean function of two variables under bi-monotonicity constraint. One future direction for research can be to develop a nonparametric approach for modeling such constrained bivariate functions that is not necessarily based on an exponential decay model. Such a model will require substantial work and will be pursued elsewhere.

\newpage

\begin{table}[ht]
\centering
\caption{DL Percentage resource available}
\label{tab:tab1}
\hspace*{-5em}
\begin{tabular}{rrrrrrrrrrrr}
\hline
 \multicolumn{11}{c}{Wickets lost} \\
\hline
 Overs Remaining     &0 & 1 & 2 & 3 & 4 & 5 & 6 & 7 & 8 & 9\\
  \hline
 50 & 100.00 & 93.40 & 85.10 & 74.90 & 62.70 & 49.00 & 34.90 & 22.00 & 11.90 & 4.70 \\ 
 45 & 95.00 & 89.10 & 81.80 & 72.50 & 61.30 & 48.40 & 34.80 & 22.00 & 11.90 & 4.70 \\ 
 40 & 89.30 & 84.20 & 77.80 & 69.60 & 59.50 & 47.60 & 34.60 & 22.00 & 11.90 & 4.70 \\ 
 35 & 82.70 & 78.50 & 73.00 & 66.00 & 57.20 & 46.40 & 34.20 & 21.90 & 11.90 & 4.70 \\ 
 30 & 75.10 & 71.80 & 67.30 & 61.60 & 54.10 & 44.70 & 33.60 & 21.80 & 11.90 & 4.70 \\ 
 25 & 66.50 & 63.90 & 60.50 & 56.00 & 50.00 & 42.20 & 32.60 & 21.60 & 11.90 & 4.70 \\ 
 20 & 56.60 & 54.80 & 52.40 & 49.10 & 44.60 & 38.60 & 30.80 & 21.20 & 11.90 & 4.70 \\ 
 15 & 45.20 & 44.10 & 42.60 & 40.50 & 37.60 & 33.50 & 27.80 & 20.20 & 11.80 & 4.70 \\ 
 10 & 32.10 & 31.60 & 30.80 & 29.80 & 28.30 & 26.10 & 22.80 & 17.90 & 11.40 & 4.70 \\ 
 5 & 17.20 & 17.00 & 16.80 & 16.50 & 16.10 & 15.40 & 14.30 & 12.50 & 9.40 & 4.60 \\ 
 1 & 3.60 & 3.60 & 3.60 & 3.60 & 3.60 & 3.50 & 3.50 & 3.40 & 3.20 & 2.50 \\ 
   \hline
\end{tabular}
\end{table}
\begin{table}[ht]
\centering
\caption{Resources Available: Non-Parametric Method}
\label{tab:tab2}
\hspace*{-5em}
\begin{tabular}{rrrrrrrrrrr}
\hline
 \multicolumn{11}{c}{Wickets lost} \\
  \hline Overs Remaining
 & 0 & 1 & 2 & 3 & 4 & 5 & 6 & 7 & 8 & 9 \\ 
  \hline
50 & 100.00 &  &  &  &  &  &  &  &  &  \\ 
  45 & 94.91 & 90.46 & 82.64 & 84.87 & 95.90 &  &  &  &  &  \\ 
  40 & 88.46 & 83.58 & 78.38 & 75.04 & 73.60 & 71.92 &  &  &  &  \\ 
  35 & 81.53 & 77.89 & 72.18 & 69.17 & 65.83 & 59.33 &  &  &  &  \\ 
  30 & 74.83 & 71.20 & 67.27 & 60.76 & 58.24 & 60.35 & 58.12 &  &  &  \\ 
  25 & 65.68 & 64.85 & 59.61 & 55.04 & 52.94 & 49.37 & 52.31 & 46.13 &  &  \\ 
  20 & 58.15 & 56.23 & 52.19 & 47.89 & 45.45 & 42.51 & 40.92 & 41.93 &  &  \\ 
  15 & 47.90 & 48.08 & 43.79 & 39.87 & 38.96 & 34.53 & 33.31 & 30.09 & 27.97 &  \\ 
  10 & 35.24 & 36.29 & 33.33 & 32.16 & 28.90 & 27.05 & 24.56 & 23.14 & 19.68 & 19.25 \\ 
  5 &  & 23.10 & 17.51 & 18.90 & 18.56 & 16.55 & 14.93 & 14.36 & 11.92 & 11.78 \\ 
  1 &  & 4.00 & 3.27 & 4.36 & 4.14 & 4.13 & 3.75 & 3.77 & 3.04 & 2.63 \\ 
   \hline
\end{tabular}
\end{table}

\begin{table}[ht]
\centering
\caption{Resources Available: Bayesian Model}
\label{tab:tab3}
\hspace*{-5em}
\begin{tabular}{rrrrrrrrrrr}
\hline
 \multicolumn{11}{c}{Wickets lost} \\
  \hline Overs Remaining
 & 0 & 1 & 2 & 3 & 4 & 5 & 6 & 7 & 8 & 9 \\ 
   \hline
50 & 100.00 & 93.64 & 85.17 & 78.34 & 72.35 & 65.95 & 62.18 & 56.32 & 45.53 & 35.59 \\ 
  45 & 95.16 & 89.50 & 81.88 & 75.51 & 70.04 & 64.05 & 60.42 & 54.91 & 44.64 & 35.09 \\ 
  40 & 89.59 & 84.65 & 77.93 & 72.06 & 67.17 & 61.65 & 58.19 & 53.10 & 43.46 & 34.38 \\ 
  35 & 83.15 & 78.96 & 73.20 & 67.89 & 63.62 & 58.64 & 55.38 & 50.77 & 41.87 & 33.39 \\ 
  30 & 75.72 & 72.30 & 67.52 & 62.83 & 59.23 & 54.85 & 51.83 & 47.76 & 39.75 & 32.01 \\ 
  25 & 67.15 & 64.48 & 60.71 & 56.70 & 53.79 & 50.08 & 47.36 & 43.89 & 36.92 & 30.06 \\ 
  20 & 57.27 & 55.33 & 52.54 & 49.26 & 47.07 & 44.07 & 41.71 & 38.92 & 33.12 & 27.33 \\ 
  15 & 45.86 & 44.60 & 42.74 & 40.25 & 38.76 & 36.52 & 34.59 & 32.51 & 28.04 & 23.50 \\ 
  10 & 32.70 & 32.02 & 30.99 & 29.32 & 28.47 & 27.01 & 25.61 & 24.27 & 21.25 & 18.12 \\ 
  5 & 17.52 & 17.28 & 16.90 & 16.06 & 15.75 & 15.05 & 14.29 & 13.66 & 12.16 & 10.58 \\ 
  1 & 3.71 & 3.68 & 3.63 & 3.47 & 3.42 & 3.29 & 3.13 & 3.01 & 2.72 & 2.41 \\ 
   \hline
\end{tabular}
\end{table}
\newpage 

\clearpage
\section*{Appendix A}
\begin{table}[ht]
\centering
\small
\caption{DL Percentage resource available }
\label{tab:tab5}
\scalebox{0.76}{
\hspace*{-5em}
\begin{tabular}{rrrrrrrrrrrr}
\hline
 \multicolumn{11}{c}{Wickets lost} \\
\hline
 Overs Remaining     &0 & 1 & 2 & 3 & 4 & 5 & 6 & 7 & 8 & 9\\
  \hline
  50 & 100.00 & 93.40 & 85.10 & 74.90 & 62.70 & 49.00 & 34.90 & 22.00 & 11.90 & 4.70 \\ 
  49 & 99.10 & 92.60 & 84.50 & 74.40 & 62.50 & 48.90 & 34.90 & 22.00 & 11.90 & 4.70 \\ 
    48 & 98.10 & 91.70 & 83.80 & 74.00 & 62.20 & 48.80 & 34.90 & 22.00 & 11.90 & 4.70 \\ 
   47 & 97.10 & 90.90 & 83.20 & 73.50 & 61.90 & 48.60 & 34.90 & 22.00 & 11.90 & 4.70 \\ 
   46 & 96.10 & 90.00 & 82.50 & 73.00 & 61.60 & 48.50 & 34.80 & 22.00 & 11.90 & 4.70 \\ 
   45 & 95.00 & 89.10 & 81.80 & 72.50 & 61.30 & 48.40 & 34.80 & 22.00 & 11.90 & 4.70 \\ 
  44 & 93.90 & 88.20 & 81.00 & 72.00 & 61.00 & 48.30 & 34.80 & 22.00 & 11.90 & 4.70 \\ 
    43 & 92.80 & 87.30 & 80.30 & 71.40 & 60.70 & 48.10 & 34.70 & 22.00 & 11.90 & 4.70 \\ 
    42 & 91.70 & 86.30 & 79.50 & 70.90 & 60.30 & 47.90 & 34.70 & 22.00 & 11.90 & 4.70 \\ 
   41 & 90.50 & 85.30 & 78.70 & 70.30 & 59.90 & 47.80 & 34.60 & 22.00 & 11.90 & 4.70 \\ 
  40 & 89.30 & 84.20 & 77.80 & 69.60 & 59.50 & 47.60 & 34.60 & 22.00 & 11.90 & 4.70 \\ 
  39 & 88.00 & 83.10 & 76.90 & 69.00 & 59.10 & 47.40 & 34.50 & 22.00 & 11.90 & 4.70 \\ 
  38 & 86.70 & 82.00 & 76.00 & 68.30 & 58.70 & 47.10 & 34.50 & 21.90 & 11.90 & 4.70 \\ 
  37 & 85.40 & 80.90 & 75.00 & 67.60 & 58.20 & 46.90 & 34.40 & 21.90 & 11.90 & 4.70 \\ 
  36 & 84.10 & 79.70 & 74.10 & 66.80 & 57.70 & 46.60 & 34.30 & 21.90 & 11.90 & 4.70 \\ 
  35 & 82.70 & 78.50 & 73.00 & 66.00 & 57.20 & 46.40 & 34.20 & 21.90 & 11.90 & 4.70 \\ 
  34 & 81.30 & 77.20 & 72.00 & 65.20 & 56.60 & 46.10 & 34.10 & 21.90 & 11.90 & 4.70 \\ 
  33 & 79.80 & 75.90 & 70.90 & 64.40 & 56.00 & 45.80 & 34.00 & 21.90 & 11.90 & 4.70 \\ 
  32 & 78.30 & 74.60 & 69.70 & 63.50 & 55.40 & 45.40 & 33.90 & 21.90 & 11.90 & 4.70 \\ 
  31 & 76.70 & 73.20 & 68.60 & 62.50 & 54.80 & 45.10 & 33.70 & 21.90 & 11.90 & 4.70 \\ 
  30 & 75.10 & 71.80 & 67.30 & 61.60 & 54.10 & 44.70 & 33.60 & 21.80 & 11.90 & 4.70 \\ 
  29 & 73.50 & 70.30 & 66.10 & 60.50 & 53.40 & 44.20 & 33.40 & 21.80 & 11.90 & 4.70 \\ 
  28 & 71.80 & 68.80 & 64.80 & 59.50 & 52.60 & 43.80 & 33.20 & 21.80 & 11.90 & 4.70 \\ 
  27 & 70.10 & 67.20 & 63.40 & 58.40 & 51.80 & 43.30 & 33.00 & 21.70 & 11.90 & 4.70 \\ 
  26 & 68.30 & 65.60 & 62.00 & 57.20 & 50.90 & 42.80 & 32.80 & 21.70 & 11.90 & 4.70 \\ 
  25 & 66.50 & 63.90 & 60.50 & 56.00 & 50.00 & 42.20 & 32.60 & 21.60 & 11.90 & 4.70 \\ 
  24 & 64.60 & 62.20 & 59.00 & 54.70 & 49.00 & 41.60 & 32.30 & 21.60 & 11.90 & 4.70 \\ 
  23 & 62.70 & 60.40 & 57.40 & 53.40 & 48.00 & 40.90 & 32.00 & 21.50 & 11.90 & 4.70 \\ 
  22 & 60.70 & 58.60 & 55.80 & 52.00 & 47.00 & 40.20 & 31.60 & 21.40 & 11.90 & 4.70 \\ 
  21 & 58.70 & 56.70 & 54.10 & 50.60 & 45.80 & 39.40 & 31.20 & 21.30 & 11.90 & 4.70 \\ 
  20 & 56.60 & 54.80 & 52.40 & 49.10 & 44.60 & 38.60 & 30.80 & 21.20 & 11.90 & 4.70 \\ 
  19 & 54.40 & 52.80 & 50.50 & 47.50 & 43.40 & 37.70 & 30.30 & 21.10 & 11.90 & 4.70 \\ 
  18 & 52.20 & 50.70 & 48.60 & 45.90 & 42.00 & 36.80 & 29.80 & 20.90 & 11.90 & 4.70 \\ 
  17 & 49.90 & 48.50 & 46.70 & 44.10 & 40.60 & 35.80 & 29.20 & 20.70 & 11.90 & 4.70 \\ 
  16 & 47.60 & 46.30 & 44.70 & 42.30 & 39.10 & 34.70 & 28.50 & 20.50 & 11.80 & 4.70 \\ 
  15 & 45.20 & 44.10 & 42.60 & 40.50 & 37.60 & 33.50 & 27.80 & 20.20 & 11.80 & 4.70 \\ 
  14 & 42.70 & 41.70 & 40.40 & 38.50 & 35.90 & 32.20 & 27.00 & 19.90 & 11.80 & 4.70 \\ 
  13 & 40.20 & 39.30 & 38.10 & 36.50 & 34.20 & 30.80 & 26.10 & 19.50 & 11.70 & 4.70 \\ 
  12 & 37.60 & 36.80 & 35.80 & 34.30 & 32.30 & 29.40 & 25.10 & 19.00 & 11.60 & 4.70 \\ 
  11 & 34.90 & 34.20 & 33.40 & 32.10 & 30.40 & 27.80 & 24.00 & 18.50 & 11.50 & 4.70 \\ 
  10 & 32.10 & 31.60 & 30.80 & 29.80 & 28.30 & 26.10 & 22.80 & 17.90 & 11.40 & 4.70 \\ 
  9 & 29.30 & 28.90 & 28.20 & 27.40 & 26.10 & 24.20 & 21.40 & 17.10 & 11.20 & 4.70 \\ 
  8 & 26.40 & 26.00 & 25.50 & 24.80 & 23.80 & 22.30 & 19.90 & 16.20 & 10.90 & 4.70 \\ 
  7 & 23.40 & 23.10 & 22.70 & 22.20 & 21.40 & 20.10 & 18.20 & 15.20 & 10.50 & 4.70 \\ 
  6 & 20.30 & 20.10 & 19.80 & 19.40 & 18.80 & 17.80 & 16.40 & 13.90 & 10.10 & 4.60 \\ 
  5 & 17.20 & 17.00 & 16.80 & 16.50 & 16.10 & 15.40 & 14.30 & 12.50 & 9.40 & 4.60 \\ 
  4 & 13.90 & 13.80 & 13.70 & 13.50 & 13.20 & 12.70 & 12.00 & 10.70 & 8.40 & 4.50 \\ 
  3 & 10.60 & 10.50 & 10.40 & 10.30 & 10.20 & 9.90 & 9.50 & 8.70 & 7.20 & 4.20 \\ 
  2 & 7.20 & 7.10 & 7.10 & 7.00 & 7.00 & 6.80 & 6.60 & 6.20 & 5.50 & 3.70 \\ 
  1 & 3.60 & 3.60 & 3.60 & 3.60 & 3.60 & 3.50 & 3.50 & 3.40 & 3.20 & 2.50 \\ 
   \hline
\end{tabular}
}
\end{table}

\begin{table}[ht]
\centering
\small
\caption{Nonparametric Percentage resource available}
\label{tab:tab6}
\scalebox{0.76}{
\hspace*{-5em}
\begin{tabular}{rrrrrrrrrrrr}
\hline
 \multicolumn{11}{c}{Wickets lost} \\
\hline
 Overs Remaining     &0 & 1 & 2 & 3 & 4 & 5 & 6 & 7 & 8 & 9\\
  \hline
  50 & 100.00 &  &  &  &  &  &  &  &  &  \\ 
  49 & 99.32 & 93.77 & 89.09 &  &  &  &  &  &  &  \\ 
  48 & 98.68 & 91.60 & 77.85 &  &  &  &  &  &  &  \\ 
  47 & 97.23 & 92.21 & 83.46 & 101.71 &  &  &  &  &  &  \\ 
  46 & 96.07 & 91.39 & 84.48 & 84.59 &  &  &  &  &  &  \\ 
  45 & 94.91 & 90.46 & 82.64 & 84.87 & 95.90 &  &  &  &  &  \\ 
  44 & 93.52 & 89.40 & 81.27 & 84.42 & 85.00 &  &  &  &  &  \\ 
  43& 92.32 & 87.66 & 81.76 & 80.03 & 84.09 &  &  &  &  &  \\ 
  42& 90.30 & 86.42 & 81.51 & 79.32 & 82.64 &  &  &  &  &  \\ 
  41 & 89.60 & 84.73 & 80.19 & 77.41 & 75.10 &  &  &  &  &  \\ 
  40 & 88.46 & 83.58 & 78.38 & 75.04 & 73.60 & 71.92 &  &  &  &  \\ 
  39 & 87.03 & 82.47 & 77.02 & 74.60 & 70.74 & 71.56 &  &  &  &  \\ 
  38 & 85.29 & 81.30 & 75.66 & 73.09 & 70.47 & 69.75 &  &  &  &  \\ 
  37 & 84.47 & 79.61 & 75.09 & 71.58 & 70.95 & 62.00 &  &  &  &  \\ 
  36 & 83.22 & 78.87 & 73.43 & 70.18 & 68.79 & 61.03 &  &  &  &  \\ 
  35 & 81.53 & 77.89 & 72.18 & 69.17 & 65.83 & 59.33 &  &  &  &  \\ 
  34 & 80.24 & 76.58 & 70.92 & 68.07 & 64.54 & 56.85 &  &  &  &  \\ 
  33 & 78.43 & 75.45 & 69.86 & 66.23 & 62.41 & 61.82 &  &  &  &  \\ 
  32 & 77.21 & 73.80 & 68.93 & 64.30 & 62.41 & 60.79 & 57.03 &  &  &  \\ 
  31 & 76.04 & 72.81 & 67.68 & 62.73 & 60.50 & 60.69 & 51.40 &  &  &  \\ 
  30 & 74.83 & 71.20 & 67.27 & 60.76 & 58.24 & 60.35 & 58.12 &  &  &  \\ 
  29 & 72.96 & 69.93 & 65.90 & 59.59 & 56.88 & 59.07 & 57.58 &  &  &  \\ 
  28 & 71.99 & 68.62 & 64.37 & 58.51 & 56.45 & 55.65 & 56.85 &  &  &  \\ 
  27 & 69.39 & 67.77 & 62.57 & 57.56 & 54.97 & 53.90 & 55.16 &  &  &  \\ 
  26 & 67.44 & 66.49 & 61.01 & 56.43 & 54.26 & 51.28 & 55.81 & 48.49 &  &  \\ 
  25 & 65.68 & 64.85 & 59.61 & 55.04 & 52.94 & 49.37 & 52.31 & 46.13 &  &  \\ 
  24 & 65.28 & 63.18 & 58.32 & 53.30 & 52.16 & 47.49 & 49.68 & 44.86 &  &  \\ 
  23 & 62.59 & 62.41 & 56.86 & 52.08 & 49.86 & 46.68 & 48.17 & 44.56 &  &  \\ 
  22 & 61.43 & 60.23 & 55.51 & 50.71 & 48.15 & 45.32 & 46.11 & 43.47 &  &  \\ 
  21 & 59.09 & 58.83 & 53.44 & 49.46 & 46.53 & 44.22 & 43.12 & 40.96 &  &  \\ 
  20 & 58.15 & 56.23 & 52.19 & 47.89 & 45.45 & 42.51 & 40.92 & 41.93 &  &  \\ 
  19 & 56.96 & 54.25 & 50.74 & 46.44 & 43.63 & 41.42 & 39.39 & 40.38 & 43.59 &  \\ 
  18 & 55.00 & 52.55 & 49.20 & 45.01 & 41.98 & 40.03 & 36.90 & 39.91 & 41.77 &  \\ 
  17 & 53.96 & 50.51 & 47.72 & 43.27 & 41.28 & 37.63 & 36.22 & 34.95 & 35.05 &  \\ 
  16 & 51.17 & 49.25 & 46.04 & 41.41 & 40.47 & 35.90 & 34.52 & 31.83 & 33.42 &  \\ 
  15 & 47.90 & 48.08 & 43.79 & 39.87 & 38.96 & 34.53 & 33.31 & 30.09 & 27.97 &  \\ 
  14 & 50.06 & 45.75 & 42.02 & 37.99 & 37.14 & 33.03 & 31.64 & 29.01 & 23.25 & 35.96 \\ 
  13 & 53.67 & 41.74 & 39.75 & 36.79 & 35.09 & 31.79 & 29.58 & 28.21 & 22.76 & 31.97 \\ 
  12 & 50.58 & 38.74 & 38.10 & 35.25 & 33.07 & 30.23 & 28.53 & 26.07 & 22.23 & 26.15 \\ 
  11 & 36.87 & 39.35 & 35.99 & 33.67 & 31.09 & 28.60 & 26.84 & 24.27 & 22.59 & 20.71 \\ 
  10 & 35.24 & 36.29 & 33.33 & 32.16 & 28.90 & 27.05 & 24.56 & 23.14 & 19.68 & 19.25 \\ 
  9 & 31.42 & 33.03 & 30.92 & 29.61 & 27.44 & 25.35 & 22.60 & 21.52 & 17.83 & 17.98 \\ 
  8 & 24.70 & 31.44 & 27.61 & 27.36 & 25.38 & 23.45 & 21.09 & 19.89 & 16.29 & 16.35 \\ 
  7 & 21.07 & 31.03 & 25.06 & 24.60 & 23.64 & 21.20 & 19.08 & 18.69 & 14.76 & 14.89 \\ 
  6 & 17.80 & 26.70 & 21.57 & 21.84 & 21.16 & 19.26 & 17.02 & 16.50 & 14.15 & 11.35 \\ 
  5 &  & 23.10 & 17.51 & 18.90 & 18.56 & 16.55 & 14.93 & 14.36 & 11.92 & 11.78 \\ 
  4 &  & 19.80 & 15.66 & 15.53 & 15.34 & 14.18 & 12.40 & 11.94 & 10.06 & 10.28 \\ 
  3 &  & 15.98 & 10.96 & 11.81 & 11.81 & 11.61 & 10.00 & 9.54 & 7.94 & 7.49 \\ 
  2 &  & 11.72 & 8.35 & 8.66 & 7.85 & 7.84 & 7.09 & 6.71 & 5.95 & 5.32 \\ 
  1 &  & 4.00 & 3.27 & 4.36 & 4.14 & 4.13 & 3.75 & 3.77 & 3.04 & 2.63 \\ 
   \hline
\end{tabular}
}
\end{table}

\begin{table}[ht]
\centering
\small
\caption{Bayesian Percentage resource available: Bayesian Model}
\label{tab:tab7}
\scalebox{0.76}{
\hspace*{-5em}
\begin{tabular}{rrrrrrrrrrrr}
\hline
 \multicolumn{11}{c}{Wickets lost} \\
\hline
 Overs Remaining     &0 & 1 & 2 & 3 & 4 & 5 & 6 & 7 & 8 & 9\\
 \hline
50 & 100.00 & 93.64 & 85.17 & 78.34 & 72.35 & 65.95 & 62.18 & 56.32 & 45.53 & 35.59 \\ 
  49 & 99.09 & 92.87 & 84.56 & 77.82 & 71.93 & 65.60 & 61.86 & 56.07 & 45.37 & 35.50 \\ 
  48 & 98.15 & 92.06 & 83.93 & 77.27 & 71.49 & 65.24 & 61.53 & 55.80 & 45.21 & 35.41 \\ 
  47 & 97.18 & 91.24 & 83.27 & 76.71 & 71.02 & 64.86 & 61.17 & 55.52 & 45.03 & 35.31 \\ 
  46 & 96.19 & 90.38 & 82.59 & 76.12 & 70.54 & 64.46 & 60.80 & 55.23 & 44.84 & 35.20 \\ 
  45 & 95.16 & 89.50 & 81.88 & 75.51 & 70.04 & 64.05 & 60.42 & 54.91 & 44.64 & 35.09 \\ 
  44 & 94.11 & 88.59 & 81.15 & 74.87 & 69.51 & 63.61 & 60.01 & 54.59 & 44.43 & 34.97 \\ 
  43 & 93.03 & 87.65 & 80.39 & 74.21 & 68.96 & 63.15 & 59.59 & 54.24 & 44.21 & 34.83 \\ 
  42 & 91.91 & 86.68 & 79.60 & 73.52 & 68.39 & 62.68 & 59.14 & 53.88 & 43.98 & 34.69 \\ 
  41 & 90.77 & 85.68 & 78.78 & 72.81 & 67.79 & 62.18 & 58.68 & 53.50 & 43.72 & 34.54 \\ 
  40 & 89.59 & 84.65 & 77.93 & 72.06 & 67.17 & 61.65 & 58.19 & 53.10 & 43.46 & 34.38 \\ 
  39 & 88.37 & 83.58 & 77.05 & 71.29 & 66.52 & 61.10 & 57.68 & 52.68 & 43.18 & 34.21 \\ 
  38 & 87.12 & 82.48 & 76.14 & 70.49 & 65.84 & 60.53 & 57.14 & 52.24 & 42.88 & 34.03 \\ 
  37 & 85.83 & 81.35 & 75.20 & 69.66 & 65.13 & 59.93 & 56.58 & 51.77 & 42.56 & 33.83 \\ 
  36 & 84.51 & 80.17 & 74.21 & 68.79 & 64.39 & 59.30 & 55.99 & 51.28 & 42.23 & 33.62 \\ 
  35 & 83.15 & 78.96 & 73.20 & 67.89 & 63.62 & 58.64 & 55.38 & 50.77 & 41.87 & 33.39 \\ 
  34 & 81.75 & 77.71 & 72.14 & 66.95 & 62.81 & 57.95 & 54.73 & 50.22 & 41.50 & 33.15 \\ 
  33 & 80.31 & 76.42 & 71.05 & 65.98 & 61.97 & 57.23 & 54.06 & 49.65 & 41.10 & 32.89 \\ 
  32 & 78.82 & 75.09 & 69.92 & 64.97 & 61.10 & 56.47 & 53.35 & 49.05 & 40.68 & 32.62 \\ 
  31 & 77.29 & 73.72 & 68.74 & 63.92 & 60.18 & 55.68 & 52.61 & 48.42 & 40.23 & 32.32 \\ 
  30 & 75.72 & 72.30 & 67.52 & 62.83 & 59.23 & 54.85 & 51.83 & 47.76 & 39.75 & 32.01 \\ 
  29 & 74.11 & 70.83 & 66.25 & 61.70 & 58.23 & 53.98 & 51.02 & 47.06 & 39.25 & 31.67 \\ 
  28 & 72.44 & 69.32 & 64.94 & 60.52 & 57.19 & 53.07 & 50.16 & 46.33 & 38.72 & 31.31 \\ 
  27 & 70.73 & 67.76 & 63.58 & 59.29 & 56.11 & 52.12 & 49.27 & 45.56 & 38.15 & 30.92 \\ 
  26 & 68.97 & 66.15 & 62.17 & 58.02 & 54.98 & 51.12 & 48.34 & 44.75 & 37.55 & 30.50 \\ 
  25 & 67.15 & 64.48 & 60.71 & 56.70 & 53.79 & 50.08 & 47.36 & 43.89 & 36.92 & 30.06 \\ 
  24 & 65.29 & 62.77 & 59.19 & 55.32 & 52.56 & 48.98 & 46.33 & 43.00 & 36.24 & 29.58 \\ 
  23 & 63.37 & 61.00 & 57.62 & 53.89 & 51.28 & 47.84 & 45.25 & 42.05 & 35.53 & 29.08 \\ 
  22 & 61.39 & 59.17 & 55.99 & 52.41 & 49.93 & 46.64 & 44.13 & 41.06 & 34.77 & 28.53 \\ 
  21 & 59.36 & 57.28 & 54.30 & 50.87 & 48.53 & 45.39 & 42.95 & 40.01 & 33.97 & 27.95 \\ 
  20 & 57.27 & 55.33 & 52.54 & 49.26 & 47.07 & 44.07 & 41.71 & 38.92 & 33.12 & 27.33 \\ 
  19 & 55.12 & 53.32 & 50.72 & 47.59 & 45.55 & 42.70 & 40.42 & 37.76 & 32.22 & 26.66 \\ 
  18 & 52.90 & 51.24 & 48.83 & 45.86 & 43.96 & 41.26 & 39.06 & 36.55 & 31.26 & 25.95 \\ 
  17 & 50.62 & 49.09 & 46.88 & 44.06 & 42.30 & 39.75 & 37.64 & 35.27 & 30.25 & 25.19 \\ 
  16 & 48.28 & 46.88 & 44.85 & 42.19 & 40.57 & 38.17 & 36.15 & 33.92 & 29.18 & 24.37 \\ 
  15 & 45.86 & 44.60 & 42.74 & 40.25 & 38.76 & 36.52 & 34.59 & 32.51 & 28.04 & 23.50 \\ 
  14 & 43.38 & 42.24 & 40.56 & 38.22 & 36.87 & 34.79 & 32.96 & 31.02 & 26.84 & 22.56 \\ 
  13 & 40.82 & 39.80 & 38.30 & 36.12 & 34.90 & 32.98 & 31.25 & 29.46 & 25.56 & 21.56 \\ 
  12 & 38.19 & 37.29 & 35.95 & 33.94 & 32.85 & 31.08 & 29.46 & 27.81 & 24.20 & 20.49 \\ 
  11 & 35.49 & 34.70 & 33.52 & 31.67 & 30.71 & 29.09 & 27.58 & 26.08 & 22.77 & 19.35 \\ 
  10 & 32.70 & 32.02 & 30.99 & 29.32 & 28.47 & 27.01 & 25.61 & 24.27 & 21.25 & 18.12 \\ 
  9 & 29.84 & 29.25 & 28.38 & 26.87 & 26.14 & 24.84 & 23.55 & 22.35 & 19.63 & 16.81 \\ 
  8 & 26.89 & 26.40 & 25.66 & 24.32 & 23.70 & 22.56 & 21.40 & 20.34 & 17.93 & 15.41 \\ 
  7 & 23.85 & 23.46 & 22.85 & 21.67 & 21.16 & 20.17 & 19.14 & 18.22 & 16.11 & 13.90 \\ 
  6 & 20.73 & 20.42 & 19.93 & 18.92 & 18.51 & 17.67 & 16.77 & 16.00 & 14.19 & 12.30 \\ 
  5 & 17.52 & 17.28 & 16.90 & 16.06 & 15.75 & 15.05 & 14.29 & 13.66 & 12.16 & 10.58 \\ 
  4 & 14.21 & 14.04 & 13.76 & 13.09 & 12.86 & 12.31 & 11.69 & 11.19 & 10.00 & 8.74 \\ 
  3 & 10.81 & 10.69 & 10.51 & 10.01 & 9.85 & 9.44 & 8.97 & 8.60 & 7.71 & 6.77 \\ 
  2 & 7.31 & 7.24 & 7.13 & 6.80 & 6.70 & 6.44 & 6.12 & 5.88 & 5.29 & 4.67 \\ 
  1 & 3.71 & 3.68 & 3.63 & 3.47 & 3.42 & 3.29 & 3.13 & 3.01 & 2.72 & 2.41 \\ 
   \hline
\end{tabular}
}
\end{table}
\clearpage

\section*{Appendix B}
\textbf{Proof of Monotonicity of Resource table under said prior specification:}\par

\noindent Consider mean function $m(u,w;\theta)=a_w(1-e^{-b_w u})$ as given in (\ref{eq:mean}) where $a_w>0$ and  $b_w>0\;\;\forall w\in\mathcal{W}$.  We show that the sequence of priors given in (\ref{eq:prior}) ensures that the following conditions as given in (\ref{eq:nec.cond}) are satisfied (with probability 1):
\begin{itemize}
\item[(i)] $m(u_1,w)< m(u_2,w)$ if $0\leq u_1\leq u_2, w \in \mathcal{W}=\{0,1,2...9\}$ 
\item[(ii)] $m(u,w)>m(u,w+1)$ $\forall u\geq0$ and $w \in \mathcal{W}$
\end{itemize}
Notice that $\frac{\partial m(u,w)}{\partial u}=a_wb_we^{-b_wu}>0$\hspace{1mm} $\forall w$. Now for fixed $w \in \mathcal{W}$ and $\forall u\geq 0$,
\begin{eqnarray}
m(u,w)>m(u,w+1)&\iff& a_w(1-e^{-b_w u})> a_{w+1}(1-e^{-b_{w+1} u})\\
&\iff& a_{w+1}e^{-b_{w+1} u}-a_{w}e^{-b_{w} u}>a_{w+1}-a_w 
\label{eq:cond1}
\end{eqnarray}
Consider the function, $g_w(u)=a_{w+1}e^{-b_{w+1} u}-a_{w}e^{-b_{w} u}$, then $g_w(0)=a_{w+1}-a_w$. Also in order for the above condition (\ref{eq:cond1}) to hold $g_w(u)>g_w(0)$ $\forall u\geq 0$, by letting $u\rightarrow \infty$ we get $g_w(\infty)=0\geq g_w(0)=a_{w+1}-a_w$ which implies $a_{w+1}\leq a_w$ $\forall w \in \mathcal{W}$. Next we consider two possible cases:
\begin{itemize}
\item[Case 1:] $b_w\geq b_{w+1}$: In this case conditions (i) and hence (ii) above are automatically satisfied if $a_{w+1}\leq a_w$.
\item[Case 2:] $b_w<b_{w+1}$: 
In this case, the derivative of $g_w$ given by $$g_w^\prime(u)=a_wb_we^{-b_wu}-a_{w+1}b_{w+1}e^{-b_{w+1}u}>0 \iff u > \frac{1}{b_{w+1}-bw}\log\left(\frac{a_{w+1}b_{w+1}}{a_wb_w}\right).$$
\end{itemize}
So if $a_{w+1}b_{w+1}\leq a_wb_w$ then $g_w^\prime(u)\geq 0\;\;\forall$ $u\geq 0$, so we have $g_w(u)>g_w(0)$ i.e the condition (\ref{eq:cond1}) holds. Therefore the following constraints on prior specifications grantees the required monotonicity.
\begin{itemize}
\item[(i)] $0\leq b_{w+1}\leq\frac{a_w b_w}{a_{w+1}}$
\item[(ii)] $a_{w+1}\leq a_w$
\end{itemize}
And our prior specifications (by using uniform distributions) maintain these constraints on the parameters to ensure the required monotonicity of the estimated resource table.
\newpage

\section*{Appendix C}
\textbf{Code Snippet for running MCMC:}\par
\begin{verbatim}
model{
for (i in 1:500)
{
R[i]~dnorm(mu[i],Tau[i])
Tau[i]<-tau*nn[i]
mu[i]<- a[w[i]] *(1-exp(-b[w[i]]*u[i]))
}
tau~dgamma(.1,.1)
sigma<-1/sqrt(tau)
a[1]~dunif(0,2000)
for(j in 2:10)
{a[j]~dunif(0,a[j-1])
}
b[1]~dunif(0,100)
for(j in 2:10)
{b[j]~ dunif(0,val[j])}

for(j in 2:10)
{val[j]<-(a[j-1]*b[j-1])/a[j]}
}
\end{verbatim}
 
In above, observed values of $\bar{R}(u,w)$ are entered for {\tt R[i]} and ``NA" are used for missing values and {\tt nn[i]} denotes the $n_{uw}$ for the number matches for which $w$ wickets were lost after $u$ overs. As there are 10 possible values for $w$ and $50$ possible values for $u$'s, we have a total of $10\times 50=500$ possibilities and entries for the resource table.

\end{document}